\documentstyle[12pt]{article}

\newcommand{\be}{\begin{equation}}
\newcommand{\ee}{\end{equation}}
\newcommand{\bear}{\begin{eqnarray}}
\newcommand{\eear}{\end{eqnarray}}

\def\la{\mathrel{\mathpalette\fun <}}

\def\fun#1#2{\lower3.6pt\vbox{\baselineskip0pt\lineskip.9pt
 \ialign{$\mathsurround=0pt#1\hfil##\hfil$\crcr#2\crcr\sim\crcr}}}

\title{Hyperon--Nucleon Final State Interaction in Kaon Photoproduction of the
Deuteron}
\author{B.O.Kerbikov\\
 Institute of Theoretical and
Experimental Physics,\\ 117218, Moscow, Russia}
 \date{}
\begin{document}

\maketitle

\vspace{2cm}
\large

\begin{abstract}
Final state hyperon-nucleon interaction in
strangeness photoproduction of the deuteron is
investigated making use of the  covariant
reaction formalism and the $P$--matrix approach to the
YN system. Remarkably
simple analytical expression for the amplitude is
obtained. Pronounced effects due to final state
interaction are predicted including the manifestation
of the 2.13 GeV resonance.
\end{abstract}

\vspace{1cm}


\newpage

\section{Introduction}
\setcounter{equation}{0} \def\theequation{1.\arabic{equation}}

Up to now most of the  information on the hyperon--nucleon (YN)
interaction has been obtained either from hypernuclei or from
$K^-d$ and $\pi^+d$ reactions. After several decades of studies
our knowledge on the YN system is still far from being complete.

Recently the interest to the
 $\Lambda N$ -- $\Sigma N$ system
has flared again in connection to the expected CEBAF
experimental results on the kaon photoproduction on the
deuterium [1]. The final state $YN$
interaction  (FSI) plays an
important role in the $\gamma d\to K^+Yn$,
$Y=\Lambda, \Sigma^0$
 reaction.
Therefore high resolution photoproduction experiments can
substantially deepen our understanding of  the YN dynamics.

The problem of FSI in
 $\gamma d \to K^{+} Y n$ reaction has been addressed by several
 authors starting from the pioneering paper by
  Renard and Renard [2,3,4]. Two novel features differ the
 present work  from the previous studies.  First,
  covariant formalism based on direct evaluation of Feynman diagrams
  is used which allows to analyse the data beyond the region of low
  energy and low momentum transfer.
  Second,
  the YN interaction is described within the
  $P$-matrix approach  which takes into
  account the subnuclear degrees of freedom and disentangle the
  dynamical singularities from kinematical threshold effects [5].
  The $P$--matrix analysis of the $YN$ interaction was presented in
  [6] (see also [7])
   and we shall use the set of parameters from [6].

    In our previous publications [8],[9] we have announced
    some preliminary results without  proving the central assertion,
   namely that the FSI effects  allow remarkably simple evaluation
  within the $P$-matrix approach. In the present  paper we give  a
  detailed derivation of this result.
   Our central result is that  YN FSI in
  strangeness photoproduction of the deuteron is quite significant
  and that $P$-matrix provides a remarkably transparent approach to
  this phenomenon.

The organization of the paper is the following. In Section 2 we
present kinematical relations relevant for the reaction and
display equations for the cross sections. Analysis of the plane
waves approximation is given in Section 3. In Section 4 we
present a summary of the   $P$-matrix approach  to the
hyperon-nucleon FSI. In Section 5 the loop diagram with  YN FSI
is calculated and expressed through the enhancement factor in
the spirit of the Migdal--Watson FSI theory.
 Numerical results
and calculations are presented in Section 6.

\section{Kinematics  and Cross-section}
\setcounter{equation}{0} \def\theequation{2.\arabic{equation}}

The reaction $\gamma d \to K^+Yn,~~ Y=\Lambda, \Sigma^0$ is a $2\to 3$
process. The corresponding kinematical relations as well as
expressions for various cross-sections and distributions are well
known (see e.g. [10]). However we find it apropriate to include a list
of basic equations in order to facilitate comparison with other
theoretical papers and to provide a framework for the presentation of
the experimental results.

We start with a general expression for the $2\to 3$ reaction
\be
d\sigma =(2\pi)^4 \delta^{(4)}(k+p_d-p_K-p_Y-p_n)\frac{|T|^2}{4I}
\frac{d^3p_K}{(2\pi)^92E_K}
\frac{d^3p_Y}{2E_Y}
\frac{d^3p_n}{2E_n},
\label{2.1}
\ee
where $k$ is the four-momentum of the initial photon, $I$ is the flux
factor which in the deuteron rest frame is equal to $kM_d$, and the
Dirac spinors are normalized according to $\bar u u=2m$. The number
of essential final state Lorentz scalar variables for 3-body final
state is 4 with the standard set being
\be
s_1=(p_K+p_Y)^2,
s_2=(p_Y+p_n)^2, t_1=(k-p_K)^2, t_2=(p_d-p_n)^2.
\label{2.2}
\ee
In addition the total energy squared $s=(k+p_d)^2=(p_K+p_Y+p_n)^2$
will also be used in equations to follow.

 The physical
 quantity we are interested
in is the double differential cross-section
 \begin{equation}
d^2\sigma\equiv   \frac{d^2 \sigma}{d\vert {\bf p}_{K} \vert
d\Omega_{K}} = \frac{1}{2^{11}\pi^{5}} \frac{{\bf p}_{K}^{2}}{k
M_{d}E_{K}} \frac{\lambda^{1/2}(s_{2}, m_{Y}^{2}, m_{n}^{2})}{s_{2}}
        \int d\Omega^{*}_{Yn} \vert T\vert ^{2} \; .
        \label{2.3}
         \end{equation}

Here $k$, ${\bf p}_{K}^{2}$, $E_{K}$ and $\Omega_{K}$ correspond to
the deuteron rest system (LS)  with $z$-axis defined by the incident
photon beam direction ${\bf k}$. The solid angle $\Omega^{*}_{Yn}$ is
defined in the $Yn$ center-of-momentum system (CMYn). The quality
$\lambda(x, y, z)$ is the standard kinematical function
 $\lambda(x, y, z) = x^{2} - 2(y+z)x + (y-z)^{2}$.
 Integration over $d\Omega^*_{Yn}$ can  be
 replaced by integration over $dt_2 ds_1$ since linear relations hold
 between $t_2$ and $\cos \theta^*$ and $s_1$ and $\cos\varphi^*$
 correspondingly.
 With such a replacement we can write $d^2\sigma$ in terms of the
 Chew-Low plot as
 \be
  \frac{d^2
\sigma}{ds_2dt_1}=\frac{1}{2^{10}\pi^4\lambda(s,0,M^2_d)}
\int dt_2 ds_1
\frac{|T(t_2, s_1)|^2}{[-\Delta_4(t_1,s_1,s_2,t_2)]^{1/2}},
\label{2.4}
\ee
where $\Delta_4$ is a $4\times 4$ symmetric Gram determinat(see [10]
for an explicit expression) which is a quadratic in any of the
variables $t_1,s_1,s_2,t_2$.  The region of integration in equation
(2.4) has to satisfy $\Delta_4\leq 0$.  For $s_2$ and $t_1$ fixed
(which corresponds to fixed $|{\bf p}_k|$ and $\Omega_K$) the
boundary of integration $\Delta_4=0$ can  be solved for $s_1$ in
terms of $t_2$:
\be
s_1^{\pm}=m^2_K+m_Y^2-\frac{1}{2t_1}\{(t_1+m^2_K)(t_1+m^2_Y-t_2)\pm
\lambda^{1/2}(t_1, m^2_K,0)\lambda^{1/2} (t_1, m_Y^2, t_2)\}.
\label{2.5}
\ee
 Expressed in terms of $s_1^\pm$, the function $\Delta_4$ reads
 \be 16\Delta_4=\lambda(s_2, t_1, M^2_d)(s_1-s_1^+)(s_1-s_1^-),
 \label{2.6}
 \ee
 Actual calculations presented below were performed making use of
(2.4) since integration over invariant  variables $t_2 $ and $s_1$
is more straightforward than integration over the angle
$\Omega^*_{Yn}$ needed in (2.3).

\section{The Plane Waves Approximation}
\setcounter{equation}{0} \def\theequation{3.\arabic{equation}}

We shall use covariant relativistic approach to calculate the
amplitude $T$ of the $\gamma d\to K^+Yn$ process. The amplitude
will be approximated by the contributions of the two diagrams,
namely the tree (pole, or plane waves) graph and the triangle
graph with $Yn$ FSI. It will be shown that within approximations
to be specified below the sum of these diagrams reduces to the
plane waves diagram multiplied by an enchancement factor
depending on $s_2$ and therefore factorizing out  of the
integral in (2.4)

  To calculate the plane-waves diagram two blocks  have be
  specified:  {\em (i)} the
  elementary photoproduction amplitude $M^{\gamma K}$ on the proton,
  and {\em (ii)} the deuteron vertex $\Gamma_d$.  The elementary
 amplitude used in the present calculation was derived from the tree
 level effective Lagrangian with the account of several resonances in
$s,t$ and $u$ chanels [11].  This amplitude has the following
decomposition over invariant terms [12]
 \begin{equation}
  M^{\gamma K}
= \overline{u}_{Y} \sum_{j=1}^{6} {\cal A}_j {\cal M}_{j}(s', t', u')
 u_{p} \; ,
 \label{3.1}
  \end{equation}
   where
   \be
   s'  =
  (k+p_{p})^{2}, ~~t'  = (k-p_{K})^{2},~~ u' = (k-p_{Y})^{2}.
  \label{3.2}
  \ee

 The decomposition of the deuteron  vertex function $\Gamma_{d}$ in
independent Lorentz structures has the form [13]

$$
\Gamma_{d}        =  \sqrt{m_{N}}
\left[ (p_{p} + p_{n})^{2} -M_{d}^{2} \right]\times
$$
\be
     \left[\varphi_{1}(t_{2}) \frac{(p_{p}-p_{n})_{\mu}}{2m_{N}^{2}}
     + \varphi_{2}(t_{2}) \frac{1}{m_{N}} \gamma_{\mu} \right] {\cal
            E}^\mu(p_{d}, \lambda).
            \label{3.3}
\ee

Here ${\cal E}^\mu(p_{d}, \lambda)$ is the polarization
4-vector of the deuteron with momentum $p_{d}$ and polarization
$\lambda$.
 The vertex (3.3)  implies that the deuteron as well as the spectator
 neuteron are on mass shell while the proton  is off its mass
 shell.In the nonrelativistic limit (but with relativistic
 corrections) the functions $\varphi_1$ and $\varphi_2$ may be
 expressed in terms of the deuteron $S$- and $D$- waves [14]
\be
\varphi_{1}      = \frac{m_{N}^{2}}{2 {\bf p}_{n}^{2}}
[-(1 +\frac{m_{N}}{2E_{n}} ) u_{D} + \frac{1}{\sqrt{2}}
(1-\frac{m_{N}}{E_{n}}) u_{S} ]  ,
\label{3.4}
\ee
\be
\varphi_{2}  = \frac{m_{N}}{2 E_{n}} (\sqrt{2} u_{S} + u_{D}) \; ,
\label{3.5}
\ee

 with $E_{n} = \sqrt{{\bf p}_{n}^{2}+m_{N}^{2}}$.
The normalization condition is the following

\begin{equation}
 \frac{m_{N}}{2 \pi^{2}}
 \int_{0}^\infty  \frac{dp_{n} {\bf p}_{n}^{2}}{E_{n}}
 \; [u_{S}^{2}(p_{n}) + u_{D}^{2}(p_{n})]  = 1 \; .
        \label{3.6}
\end{equation}

    Now we can
 write the following expression for the tree diagram

\begin{equation}
    T^{(t)} = \overline{u}_{Y}
    \left\{ \left (\sum_{j=1}^{6} {\cal A}_{j} {\cal M}_{j} (s', t', u')
    \right )
    S(p_{p}) \Gamma_{d} \right\} u_{n}^c \; ,
        \label{3.7}
\end{equation}

\noindent where $S(p_p)$ is the proton  propagator and $u_{n}^c$ is a
charge conjugated neutron spinor.
 The deuteron  vertex in (3.7) may be substituted by the relativistic
 deuteron wave  function according to [15]
 \be
 \psi_d=[2(2\pi)^3 M_d]^{1/2} S(p_p)
\Gamma_d.
\label{3.8}
\ee
 Then (3.7) can be rewritten in the following   schematic form
\be
T^{(t)}= [2(2\pi)^3M_d]^{1/2}M^{\gamma K}\psi_d,
\label{3.9}
\ee
where $\psi_d$ is the relativistic deuteron wave function
discussed at length in [15], and
where summation over magnetic quantum numbers is tacitly assumed.

In the numerical calculations of the plane waves diagram use has
been made of the Gross relativistic deuteron wave function [15].
Finally we mention that nonrelativistic reduction of Eqs. (2.3),
(2.4) and (3.9) is straightforward and yields the standard
impulse approximation
\begin{equation}
\frac{        d^{2}\sigma}{d|{\bf p}_K|d\Omega_K} =
        \frac{{\bf p}_{K}^{2} \vert {\bf p}_{Y}^{*} \vert}{64\pi^{2}k
        E_{K} \sqrt{s_{2}}}
    \int d\Omega^{*}_{Yn} \; \vert M^{\gamma K} \vert^{2}
    \vert \psi_{d}^{nr} \vert^{2} \; ,
        \label{3.10}
\end{equation}
where the nonrelativistic deuteron wave function normalization
condition is
\begin{equation}
    \int d^{3}p_{n}  \vert \psi_{d}^{nr}(p_{n}) \vert^{2} = 1 \;.
        \label{3.11}
\end{equation}

\section{Hyperon-Nucleon interaction in the $P$-matrix approach}
\setcounter{equation}{0} \def\theequation{4.\arabic{equation}}

Before we consider the triangle diagram with $Yn$ rescattering it is
worth to recapitulate the main results of the $P$-matrix approach to
hyperon-nucleon interaction [6]. Since FSI is essential only at low
relative momenta of the $Yn$ pair the nonrelativistic threatment is
quite legitimate.

First we recall the interpretation of the $P$-matrix in terms of the
underlying coupled-channel quark-hadron potential [16]. Namely, the
Jaffe-Low  $P$-matrix description is equivalent to the coupling
between hadron and  quark channels via the nonlocal energy dependent
potential of the form
 \be
 V_{\rm hqh} =\sum_n\frac{f_n(r)f_n(r')}{E-E_n},
 \label{4.1}
 \ee
 where $E_n$ are the energies of the six-quark "primitives" [17,5]
 ($P$-matrix poles), and the form factors $f_n(r)$ are given by
 \be
 f_n(r)=\frac{c_n}{b\sqrt{4\pi}}\delta (r-b).
 \label{4.2}
 \ee
 Here $b$ is
 the bag radius [17,16,5] and $ |c_n|^2=\lambda_n^2/2\mu_{Yn}$, where
  $\lambda_n$ are the residues  of the
 $P$-matrix [17,16,5]
  \be
   P=P_0+\sum_n\frac{\lambda_n^2}{E-E_n}.
   \label{4.3}
     \ee
      As it
 was shown in [6]   a single primitive at $E_n=2.34$ GeV is
 sufficient for a high quality description of the  existing YN
 experimental data, so that the subscript $n$ will be omitted.
 To avoid lengthy expressions (4.3) is written for the case of
 a single hadron channel, namely     $\Lambda N$, inclusion of the
  $\Lambda N-
 \Sigma N$ conversion is straightforward [6] and the final results with
 the account of the $\Sigma N$ channel will be presented below.

 The scattering problem corresponding to the interaction (4.1),(4.2)
 can be easily solved. In particular for the momentum
 space half-of-shell amplitude
 $F(p,p';E'),E'=p^{'2}/2\mu_{Yn}\neq p^2/2\mu_{Yn}$
 and for the Jost function $f(-p')$ one gets
 \be
 F(p,p';E')=
 -\frac{2\pi}{\mu}
 \lambda^2 b^2\frac{\sin pb}{pb}d^{-1}(E')\frac{\sin
 p'b}{p'b},
 \label{4.4}
  \ee
   \be
    d(E')=E'-E_n+\frac
    {\lambda^2}{p'}
 e^{ip' b} \sin p'b,
 \label{4.5}
   \ee
   \be
   f^{-1}(-p')=\frac{E'-E_n}{E'-E_n+\frac
    {\lambda^2}{p'}
 e^{ip' b} \sin p'b}.
 \label{4.6}
   \ee
   These results will be used in the next  section devoted to the
   calculation of the triangle diagram with $Yn$ FSI.

   \section{Triangle Diagram }
\setcounter{equation}{0} \def\theequation{5.\arabic{equation}}

   Next we consider the triangle (loop) diagram with $Yn$ FSI. The
   general expression for this graph reads
\begin{equation}
    T^{(l)} = \int \frac{d^{4} p_{n}}{(2\pi)^{4}} \;
    \overline{u}_{Y}(p'_{Y})
    \left\{ \left( \sum_{j=1}^{6} {\cal A}_{j} {\cal M}_{j} \right)
         S(p_{p})  \Gamma_{d} C S(p_{n}) T_{Yn} S(p_{Y}) \right\}
         \overline{u}_n(p'_{n}) \; .
        \label{5.1}
\end{equation}

Here $C$ is the charge conjugation matrix, $T_{\rm YN}$ is a
hyperon-nucleon vertex, this vertex being ``dressed'' by the
corresponding spinors constitutes the hyperon-nucleon amplitude
$F$ (see (4.4)).

The comprehensive treatment of the triangle  diagram is  a purpose of
the  forthcoming  publication while here we resort to
 some  approximations with the aim to expose the effects of the
FSI.  Only positive frequency components are kept in the propagators
$S(p_n)$ and $S(p_Y)$ in (5.1), while the propagator $S(p_p)$
together with $\Gamma_d$ is lumped into the relativistic deuteron
wave function $\psi_d$ [15]. Then integration over the time component
$dp^0_n$is performed. Thus we arrive at the following expression for
$T^{(l)}$

\begin{equation}
T^{(l)} = 2\mu [{(2\pi)^{3} 2 M_{d}} ]^{1/2}
\int \frac{d{\bf p}_n}{(2\pi)^{3}}
\frac{M^{\gamma K} \psi_{d}(p_n) F (p,p';E')}%
{{\bf p}^{2} - {\bf p}^{\prime 2} -i0} \; ,
        \label{5.2}
\end{equation}

\noindent where ${\bf p}$ and ${\bf p' }$ are the YN c.m.
momenta  before and  after the    FSI, ${\bf p}={\bf
p}_n-\frac{1}{2} ({\bf k} -{\bf p}_K)$. The quality $F(p,p';E')$
is the half-off-shell Yn amplitude  given by (4.4)-(4.5).
 The
arguments of the elementary photoproduction  amplitude
$M^{\gamma K}$ in (5.2)  are specified  by (3.1)-(3.2)
 but in the kinematical region where FSI is essential $M^{\gamma K}$
 can be considered as point-like and hence factored out of the
integral (5.2) at the values of the arguments fixed by the energies
and momenta of the initial and final particles (i.e.  at the values
of the arguments corresponding to the plane-wave diagram).  Thus
(5.2) takes the form
\be T^{(l)} =-4\pi \lambda^2b^2[(2\pi)^32M_d]^{1/2}d^{-1} (E')
\frac{\sin p' b}{p'b} M^{\gamma K}  J(p'),
\label{5.3}
 \ee
  where
   \be
    J(p')=\int \frac{d{\bf p}_n}{(2\pi)^3}
\frac{\sin pb}{pb}\frac{\psi_d(p_n)}{{\bf p}^2-{\bf p'}^2-i0}.
\label{5.4}
 \ee
  In the complex $p$ plane integral (5.4) picks up
 contributions from the poles of the propagator and from the
 singularities of the deuteron wave function.  The dominant
 contribution stems from the poles of the propagator.  Numerical
 calculations using realistic  deuteron wave function will be
 presented elsewhere. Here we present simple
 arguments  illustrating  the above conclusion. Consider zero-range
 deuteron wave function which in the
 momentum space has the form
  \be
  \psi_d(p_n)=\frac{\sqrt{\alpha}}{\pi({\bf p^2_n}+\alpha^2)},
  \label{5.5}
  \ee
  where $\alpha=\sqrt{m_n|E_B|}\simeq 45.7$ MeV.  Recalling that
  ${\bf p}_n={\bf p} +{\bf Q}/2$, ${\bf Q}={\bf k}-{\bf p}_K$, and
  performing in (5.4) the angular integration, one gets
  $$
  J(p') =\frac{\sqrt{\alpha}}{\pi}\int^\infty_0\frac{dpp^2}{(2\pi)^3}
  \frac{\sin pb}{pb} ({\bf p}^2-{\bf
  p}^{\prime 2}-i0)^{-1}\frac{2\pi}{pQ} \ln \left\vert\frac
  {(p+\frac{Q}{2})^2+\alpha^2}{
  (p-\frac{Q}{2})^2+\alpha^2}\right\vert\simeq
  $$
  \be
  \simeq \frac{16\sqrt{\alpha}}{Q} \int^\infty_0 \frac{dp
  p^2}{(2\pi)^3}\frac{ \sin pb}{pb} ({\bf p}^2-{\bf p}^{\prime
  2}-i0))^{-1} (1-\frac{4\alpha^2}{Q^2}).
  \label{5.6}
  \ee
  In making the approximation (5.6) use was made of the fact that FSI
  is effective at $pb\la 1$ while the typical momentum transfer is
  $Q\gg 1/b$ and $Q\gg \alpha$. Formally this is reflected in the
  fact that the integral $J(p')$ is damped at high values of $p$ due
  to the formfactor $\sin pb/pb$. the remaining integral (5.5) is
  trivial and yields
  \be
  J(p')=(\frac{4\sqrt{\alpha}}{\pi Q^2})\frac{e^{ip'b}}{4\pi b}
  (1-\frac{4\alpha^2}{Q^2})\simeq \psi_d(p_n) \frac{e^{ip'b}}{4\pi b}
  (1-\frac{4\alpha^2}{Q^2}).
  \label{5.7}
  \ee
  Then using (5.3) and (4.6) one gets for the loop diagram
  \be
  T^{(l)}= T^{(t)} [-1+\frac{1}{f(-p')}(1-\frac{4\alpha^2 c}{Q^2})].
  \label{5.8}
  \ee
  Here $T^{(t)}$ is the plane-waves amplitude (3.9), $ f(-p')$ is the
  Jost function given by (4.6) and
  $$
  c=\frac{\lambda^2 b}{E_n-E} \frac{\sin p'b}{p'b}.
  $$
  Within few percent accuracy the term $4\alpha^2 c^2/Q^2$ can be
  neglected since $Q^2\gg \alpha^2$ and $c\ll 1$. The last inequality
  is based on the set of parameters fitted to the YN interaction [6],
  namely $\lambda^2=0.014$ GeV$^2$ and $E_n=2.34$ GeV. For this set
  of parameters $c\la 0.2$ in the  region where FSI is important.

  Therefore for the sum of the plane waves and loop diagrams one gets
  \be
  T^{(t)}+T^{(l)}\simeq T^{(t)}/f(-p')
  \label{5.9}
  \ee
  Thus we arrived at a remarkably simple and physically transparent
  result in the spirit of the Migdal-Watson FSI theory.  This result
  might seem trivial since it can be immediately obtained by
  factoring out the  deuteron wave function\footnote{At the value of
  $p_n$ corresponding to the tree diagram with the same kinematical
  variables of the initial and final particles.} out of the integral
  (5.4). However apriori such an approximation does not look reliable
  and has to  be justified making us of the amplitude corresponding
  to the $P$-matrix as it was done above.  Analogous derivation with
  the realistic deuteron wave function will be presented in the
  forthcoming publication. Inclusion of the $\Lambda N-\Sigma N$
  conversion in the $P$-matrix approach is very simple [6]. Namely,
  in (\ref{5.9}) the two-channel Jost function has to be used. It is given
  by
  \be
  F^{-1}
  (-p'_1,-p'_2)=\frac{E'-E_n}{E'-E_n+\frac{\lambda^2_1}{P^0_1-ip'_1}+
  \frac{\lambda^2_2}{P^0_2-ip'_2}},
  \label{5.10}
  \ee
  where the indices 1,2 correspond to $\Lambda N$ and $\Sigma N$
  channels and the values of the  parameters were fitted in [6]. Then
  the final expression for the amplitude  corresponding to the sum of
  the plane waves and loop diagrams with couple-channel FSI included
  reads
  \be
  T^{(t)}+T^{(l)}\simeq T^{(t)}/F(-p'_1,-p'_2).
  \label{5.11}
  \ee

  Thus using simple arguments we have splitted the  description of
  the $\gamma d\to K^+Yn$ reaction into two blocks: (i) the covariant
  calculation of the plane waves diagram with relativistic deuteron
  wave function and relativistic elementary photoproduction
  amplitude, and (ii) the calculation of the $P$-matrix enhancement
  factor with a  set of parameters fitted to the existing data on
  $YN$ interaction. The accuracy of this factorization is of the
  order of few percent which can be seen from model considerations
  presented above and will be confirmed by numerical calculations in
  the forthcoming publication.

  \section{Results and Conclusions}
\setcounter{equation}{0} \def\theequation{6.\arabic{equation}}

  Here we present the results of the calculations performed according
  to Eqs. (2.4), (3.9), (\ref{5.10}) and (\ref{5.11}). In Fig. 1 we plot the
  double differential cross-section (2.3) at $|{\bf p}_k|=1.091$ GeV
  and $\Theta_K=0^0$ in a rather wide range of the photon energy
  covering the $\Sigma N$ threshold region with nearby 2.13 GeV
  resonance. The behavior of the cross section is very instructive.
Consider first the dased line corresponding to the plane waves
approximation. At this particular kinematics $(P_K=1.091$ GeV,
$\Theta_{\gamma K}=0^o$) the threshold of $\Lambda n$ production
is at $E_\gamma=1.4.$ GeV and according to  (3.10) the
 factors that governs the near threshold cross section
behavior are the phase space and the momentum dependence of the
deuteron wave function. The sigma-nucleon threshold in Fig.1
corresponds to $E_\gamma=1.5 $ GeV. The $\Sigma N$ cusp in the
cross section is practically unvisible. This fact was known long
ago -- see e.g. [18]. It is a typical situation since "In the
experimental cross sections of nuclear reactions the cusps are
rarely seen" [19]. To see a cusp one should accurately take into
account FSI and moreover in order to be pronounced the cusp
should be enhanced by nearby pole like 2.13 GeV resonance close
to $\Sigma N$ threshold. Next we turn to the solid line in Fig.
1 with FSI included. The $\Lambda N$ FSI interaction causes the
cross section to rise more rapidly close to threshold as was
noticed before  in [3]. Then one observes a very clear signal at
the $\Sigma N$ threshold ($E_\gamma = 1.49$) GeV caused by the
pole at 2.13 GeV. We  remind that according to the
  analysis of [6,7] the 2.13 GeV resonance is not a genuine six-quark
  state but a $P$-matrix partner of the deuteron. Our prediction that
  this resonance can be investigated in strangeness photoproduction
  reaction is at odds with the conclusions made in [3] that FSI is
  important only close to $\Lambda n$ threshold and is completely
  insignificant at higher energies. The attribution of the pole
  corresponding to the 2.13 GeV resonance to a particular sheet of the
  multi-sheeted Riemann surface is a long-standing problem [6] and
  our hope is that photoproduction experiments can shed a new light
  on it.
Concerning the structure close to $\Sigma N$ threshold another
remark has to be made. Namely, one may ask whether such
structure can be generated simply by the triangle graph with
constant $\Lambda N -\Sigma N$ conversion matrix element. This
question has been addressed in the literature for the case of a
similar $K^-d\to \pi^-\Lambda p$ reaction [20,21], and it was
shown that the singularity of the  triangle diagram by itself
can not explain the observed structure.
  Finally a remark is due on another approximation tacitly made in
  the present work. This concerns hadronic corrections like
  pion-exchange to the basic  $P$-matrix YN interaction as well as
  corrections due to nonorthogonality of the six-quark and YN
  ststes [16]. As was shown in [6] the present experimental data on
  YN interaction are not sufficient to feel these corrections. From
  the theoretical side such corrections modify the Jost function
  given by (4.6) and (\ref{5.11}), in particular they smear the zero of the
  inverse Jost function at $E=E_n$.

 The author would like to thank  V.A.Karmanov for many fruitful
 discussions  and suggestions. Help in numerical calculations from
 N.O.Agasian is gratefully
  acknowledged. This work was possible
 due to stimulating contacts with C.Fayard, G.H.Lamot, F.Rouvier  and
 B.Saghai.  Hospitality and financial support from the University
 Claude Bernard and DAPNIA (Saclay)
  are gratefully acknowledged.  
 \newpage

 \begin{center}
 {\bf Figure Caption}
 \end{center}

 Fig.1
 The double differential cross section as a function of the photon
 energy for $p_K=1.091 $ GeV, $\Theta_{\gamma K}=0^o$. Dashed line
 corresponds to  the plane waves approximation, solid line -- with
 $Yn$ FSI included.

           \end{document}